# ViewpointS: towards a Collective Brain

Philippe Lemoisson[1, 2] and Stefano A. Cerri[3]

[1] CIRAD, UMR TETIS, F-34398 Montpellier, France
[2] TETIS, Univ Montpellier, AgroParisTech, CIRAD, CNRS, IRSTEA, Montpellier, France
[3] LIRMM, Univ Montpellier, CNRS, 161 Rue Ada, 34095 Montpellier, France

`philippe.lemoisson@cirad.fr;cerri@lirmm.fr`

**Abstract.** Tracing knowledge acquisition and linking learning events to interaction between peers is a major challenge of our times. We have conceived, designed and evaluated a new paradigm for constructing and using collective knowledge by Web interactions that we called ViewpointS. By exploiting the similarity with Edelman's Theory of Neuronal Group Selection (TNGS), we conjecture that it may be metaphorically considered a Collective Brain, especially effective in the case of trans-disciplinary representations. Far from being without doubts, in the paper we present the reasons (and the limits) of our proposal that aims to become a useful integrating tool for future quantitative explorations of individual as well as collective learning at different degrees of granularity. We are therefore challenging each of the current approaches: the logical one in the semantic Web, the statistical one in mining and deep learning, the social one in recommender systems based on authority and trust; not in each of their own preferred field of operation, rather in their integration weaknesses far from the holistic and dynamic behavior of the human brain.

## 1 Introduction

On one side, today's research on the human brain allows us to visualize and trace the activity along the beams connecting the neural maps. When publishing the Theory of Neuronal Group Selection (TNGS) more than thirty years ago, G. M. Edelman emphasized the observation/action loop and the social interactions loop. Both loops continuously evolve the beams connecting the neural maps under the supervision of our homeostatic internal systems also called system of values, and generate learning. On the other side, we live a digital revolution where the Web plays an increasing role in the collective construction of knowledge; this happens through the semantic Web and its ontologies, via the indexing and mining techniques of the search engines and via the social Web and its recommender systems based on authority and trust.

The goal of our approach is twofold: i) to exploit the metaphor of the brain for improving the collective construction of knowledge and ii) to better exploit our digital traces in order to refine the understanding of our learning processes. We have de-signed and prototyped a Knowledge Graph where resources (human or artificial agents, documents and descriptors) are dynamically interlinked by beams of digital connections called viewpoints (human viewpoints or artificial viewpoints issued from algorithms). We-as-agents endlessly exploit and update this graph, so that by similarity with the TNGS, we conjecture that it may be metaphorically considered a Collective Brain evolving under the supervision of all our individual systems of values.

In section 2, we present a schematic view of the biological bases of cognition, starting by the "three worlds" of K. Popper (1978) where the interaction between minds can be studied. We then re-visit the TNGS and the role played by our system of values (internal drives, instinct, intentionality…). We finally illustrate "learning through interaction" as exposed by D. Laurillard and J. Piaget.

In section 3, we explore the collective construction of knowledge in the Web paradigm, assuming that a large proportion of the traces we produce and consume today are digital ones. We distinguish three paradigms, respectively governed by logics, by statistics and by authority and trust. Thus our challenge is to integrate these paradigms and describe how individual systems of values participate to learning events.

Section 4 is dedicated to the ViewpointS approach, as a candidate for answering the challenge. The metaphor of "neural maps interconnected by beams of neurons" led to the design of a graph of "knowledge resources interconnected by beams of viewpoints", where each agent can exploit the traces of others and react to them by adding new traces. As a result, the combination of all individual "system of values" regulates the evolution of knowledge. We conjecture that it may be metaphorically considered a Collective Brain.

We conclude by recapitulating our proposal which has the limits inherent to any integrator: we are not yet sure if the collective knowledge emerging from our proposed Collective Brain will perform competitively with the existing separate paradigms respectively governed by logics, by statistics and by authority and trust. Nevertheless, if our proposal does not ensure scientific discovery about learning, we hope it represents a progress toward its comprehension.

## 2   A Schematic View of the Biological Bases of Cognition

In this section, we start by adopting a well-known philosophical position where the questions of cognition and interaction can be addressed. Then we draft a schematic view of the lessons learned from Edelman about the biological mechanisms supporting cognition, and finally we use this representation within D. Laurillard's conversational learning scenario in order to test it against the question of knowledge acquisition through interaction.

### 2.1   The Three Worlds

To start with our analysis about minds in interaction, we need some philosophical default position; "the three worlds" of K. Popper [1] provides a relevant framework. Such a framework had already found an expression in the semantic triangle of Odgen and Richards [2]. The strong interconnection of the three worlds is developed in [3] where J. Searle explains how the interpretation of repeated collective experiences by individuals bears the emergence of an institutional reality founded on the use of language. In the following, we shall refer to the three worlds as $W_1$, $W_2$ and $W_3$, with the following definitions:

$W_1$ is the bio-physical world where objects and events exist independently from us, from our perceptions, our thoughts and our languages. Causal relations, insofar we are not directly implied by some event, are also considered independent from us.

$W_2$ is the internal world of subjectivity, where the perception of objects and events of $W_1$ leave traces in memory that are combined in order to participate to the construction of our own knowledge, our consciousness about the world, where intentions appear and the emotions that will be the trigger for our actions.

$W_3$ is the world of the cultures and languages, made of interpretable traces: signs, symbols, rules of behavior and rules for representing objects and events of $W_1$. $W_3$ is the support of communication among individuals. Within $W_3$, we find all specialized languages of the scientific disciplines, as well as the language of emotions and feelings, for instance represented by smileys. Digital images such as satellite images or scanned documents are also part of $W_3$.

$W_1$ is where it happens, $W_3$ is where we can communicate about what happens, and $W_2$ is where the links and the learning events are. For this reason, we are going to pay special attention to the internal world $W_2$.

## 2.2 The Internal World of the Mind

This section pays a heavy tribute to the work of G. M. Edelman [4], [5], founder of the Theory of Neuronal Group Selection (TNGS), and one the firsts to emphasize that the brain is not a computer, but a highly dynamic, distributed and complex system, maybe the most complex "object" of the known universe. There is neither correlation between our personality and the shape of our skull (despite the teachings of phrenology), nor localized coding of information; no autopsy will ever reveal any single chunk of knowledge available in the brain.

According to the TNGS, every brain is twice unique: first because its cellular organization results from the laws of morphogenesis. Most important, however, is Edelman's second reason for the brain uniqueness: the brain is a set of "neural maps" continuously selected according to the individual's experiences. These cards, or adaptive functional units, are bi-directionally linked one-another by a fundamental integrating mechanism: the "re-entry". This crucial hypothesis allows a functional integration requiring neither any "super-card" nor any "supervising program": the neural maps are like "musicians of an orchestra linked one-another by wires in the absence of a unique conductor". The bi-directional re-entry links are the result of a selective synaptic reinforcement among groups of neurons; similarly: the cards result from a synaptic reinforcement internal to each group of neurons composing them. These reinforcements are triggered and managed by the homeostatic internal systems, also called "system of values" of each individual.

Fig. 1 (/left part of the figure) shows an observation-action loop that highlights several brain cards re-entering when grasping an apple. This type of loops originates

the perceptual categorization event, common to all highly evolved organisms, decisive for adapting the behavior to the likelihood of benefits or dangers.

In humans as well as in some higher mammals, there is a second level of categorization, supported by cards situated in the temporal, frontal and parietal areas. Beyond the immediate cartography of the world, humans may shape some durable concepts (conceptual categorization) that consider the past and/or the future.

Finally, the human brain parts specialized in language (the Wernicke and Broca areas) play a major role in the emergence of a consciousness of a higher level, enabling the human subject to "map" his-her own experience and study him-herself.

The basic principles of the TNGS (selective reinforcement and re-entry according to the advantages offered to the subject) potentially explain any learning process, from simple memorization to skill acquisition and knowledge acquisition. All these processes are regulated by our system of values.

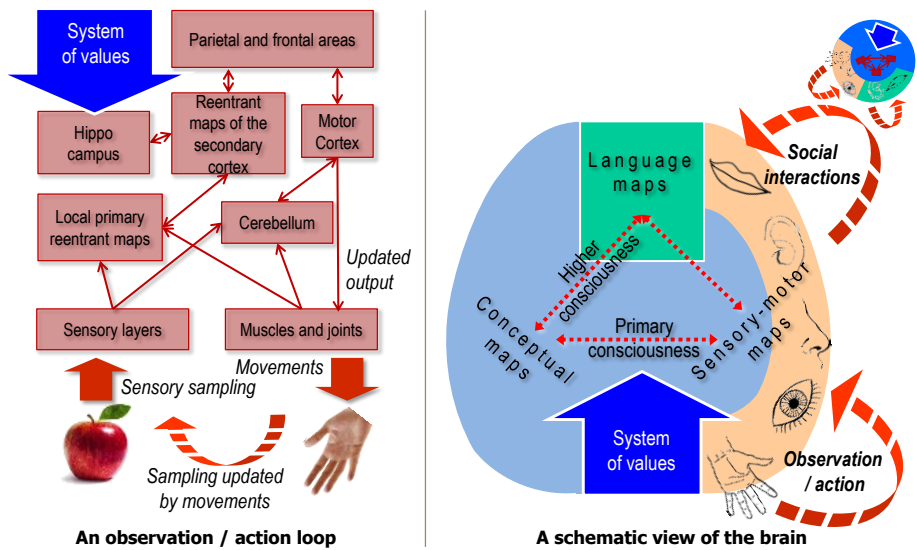

**Fig. 1.** The brain according to the TNGS of G.M. Edelman: a complex network of re-entrant maps in interaction loops with the world

A kernel element of the TNGS is quite relevant for us: knowledge is supported by a "physiological complex and adaptive network of neural maps"; the metaphor of "knowledge graph" is therefore justified. It induces to search for a topology allowing to define distances and proximity, like it was conjectured by the "zone of proximal development" of Vygotsky [6], [7]; such a topology will be presented in chapter 4.

### 2.3 Minds in Social Interaction

According to the two loops at the right of Fig. 1, we always learn through interaction: observation/action versus social interaction. These two loops clearly appear in D. Laurillard's work [8] when analyzing the acquisition of knowledge in higher education. In her scenario, a student and his teacher simultaneously experiment and discuss.

In [9] we extended this scenario to interactions within a group of peers co-constructing a representation of a shared territory.

In this multi-peers scenario, interactions occur at two levels: i) peers act in the shared territory (in $W_1$/ objects and events) and ii) peers exchange personal views of the shared territory (in $W_3$/ language). Doing so, the assimilation / adaptation processes described by J. Piaget in [10] are activated (in $W_2$/ mind), which can be interpreted in terms of series of re-entry loops according to the TNGS. In [11] we proposed a roadmap for understanding and forecasting cognitive and emotional events linked to serendipitous learning.

As a consequence of all above processes, inner views tend to synchronize and yield a shared representation. We propose that what happens on the Web is a generalization of this prototypical scenario, and can be called collective knowledge acquisition. Our approach aims at tracing it; this will be exemplified in chapter 4.

## 3 Humans in Web Interaction

The change in our lives that we have been experiencing since when Internet has gained a significant place, often called the digital revolution, has been theoretically addressed by several authors, among which S. Vial [12] and D. Cardon [13] who respectively provide a philosophical and a sociological approach. This revolution has suggested a significant hope: Internet as a space of shared knowledge able to bring in new levels of understanding in the sense given by Gruber in [14]. Internet is a support for a huge set of digital traces interpretable by humans but also by machines; if we refer to the conceptual framework above, it is part of $W_3$/language.

This space is far from being homogeneous however, and the approaches to co-build shared knowledge are multiples; hereafter, we consider three paradigms.

The first paradigm is governed by the *logical* evidence: we usually call semantic Web this logically structured part of Internet where humans interact with databases encoding the knowledge of experts according to consensual conceptual schemes such as ontologies. This allows logical responses (provided by reliable algorithms) to correctly formulated questions (and only to questions with such a property). But there are problems and limits. Firstly, ontologies only represent a fragment of the reality, and the consensus they reflect is necessarily local and temporary. Secondly, formal query languages assume a closed world – what is rarely the case. Thirdly, formal query languages require a learning effort in order to be used properly. Finally, interconnecting ontologies and supporting their evolution with time in a rapidly changing world are very heavy and costly processes. Various approaches based on automatic alignment [15] or machine learning [16] exist, but the task is endless since each ontology's evolution is domain-dependent.

The second paradigm is governed by the *statistical* evidence. The issue is to exploit techniques of data mining, i.e.: scan without too many assumptions a corpus, also called data set, of tweets, sequences, clicks, documents, … and detect regularities, frequencies, co-occurrences of items or terms. In other words: to feed suitable algorithms with the big data in order to reveal regularities. By reducing the complexity of the digital world $W_3$, the mining algorithms make it readable for humans. However, the simplicity of these descriptions must pay a price to the expressiveness or even to the effectiveness: we are just shown the surface and not the depth, the "meaning". Today, a simple question with three independent keywords on Google may give very disappointing results. What is even worse, inferential statistics – the only one allowing us to take significant decisions - require selecting the data according to a predefined goal, and this remains hidden from the user. Interpreting the results in order to build chunks of science is therefore heavily biased.

The third paradigm is based on authority and trust, which rely on *emotions* and *feelings*. The algorithms of the social Web provide information search and recommendations by graph analysis of the various personal, subjective and spontaneous light traces such as 'likes', 'bookmarks' and 'tweets'. They clearly operate in an open world; the limits are firstly the impossibility to logically assess the quality of responses and secondly the absence of guarantee concerning their stability along time.

Coming back to the dream of Gruber and many others to fuse the three paradigms, a first attempt is the semantic Web project [17] which somehow aims at subsuming them within the *logical* one; after a first wave of enthusiasm it seems that the limits listed above resist, even if they are daily pushed forward. The ViewpointS approach discussed in chapter 4 aims to offer a potential step forward in the direction of subsuming the three paradigms within the third one i.e., building up upon *trust* towards 'peers', would they be humans, databases or mining algorithms.

## 4   The ViewpointS Approach Discussed and Exemplified

This section first briefly recalls the ViewpointS framework and formalism for building collective knowledge in the metaphor of the brain - a detailed description can be found in [18], [19] - and then illustrates them through an imaginary case.

In the ViewpointS approach, the "neural maps interconnected by beams of neurons" are transposed into a graph of "knowledge resources (agents, documents, topics) interconnected by beams of viewpoints". The "systems of values" of the agents influence not only the viewpoints they emit, but also the way they interpret the graph.

We call *knowledge resource*s all the resources contributing to knowledge: agents, documents and topics. We call *viewpoints* the links between *knowledge resource*s. Each *viewpoint* is a subjective connection established by an agent (Human or Artificial) between two *knowledge resources*; the *viewpoint* $(a_1, \{r_2, r_3\}, \theta, \tau)$ stands for: the agent $a_1$ believes at time $\tau$ that $r_2$ and $r_3$ are related according to the emotion carried by $\theta$. We call Knowledge Graph the bipartite graph consisting of *knowledge resources* and *viewpoints*. Given two *knowledge resources*, the aggregation of the beam of all connections (*viewpoints*) linking them can be quantified and interpreted as a proximity. We call *perspective* the set of rules implementing this quantification by evaluating each viewpoint and then aggregating all these evaluations into a single value. The *perspective* is tuned by the "consumer" of the information, not by third a part "producer" such as Google or Amazon algorithms; each time an agent wishes to exploit the knowledge of the community, he does so through his own subjective *perspective* which acts as an interpreter.

Tuning a perspective may for instance consist in giving priority to trustworthy *agents*, or to the most recent *viewpoints*, or to the *viewpoints* issued from the logical paradigm. This clear separation between the storing of the traces (the *viewpoints*) and their subjective interpretation (through a *perspective*) protects the human agents involved in sharing knowledge against the intrusion of third-part algorithms reifying external system of values, such as those aiming at invading our psyche, influencing our actions [20], or even computing bankable profiles exploitable by brands or opinion-makers [21]. Adopting a *perspective* yields a tailored *knowledge map* where distances can be computed between knowledge resources, i.e. where the semantics emerge both from the topology of the *knowledge graph* and from our own system of values expressed by the tuned perspective.

The shared semantics emerge from the dynamics of the observation/action loops. Agents browse the shared knowledge through the *perspectives* they adopt (observation), and reversely update the graph by adding new *viewpoints* expressing their feedback (action). Along these exploitation/feedback cycles, shared knowledge is continuously elicited against the systems of values of the agents in a selection process.

To illustrate this, we develop below an imaginary case where learners have to select resources inside an Intelligent Tutoring System (ITS) to which a Knowledge Graph is associated. They wish to learn about the topic 'apple' and from step1 to step4 the learners adopt a 'neutral' perspective which puts in balance all types of viewpoints (issued from the logical or mining paradigms, or from the emotions and feelings of the learners). However at step5 where B chooses a perspective discarding his own viewpoints in order to discover new sources of knowledge.

Step1 illustrates the initial state of the knowledge. A, B and C are co-learners in the ITS (linked as such within the logical paradigm); the big arrows within the icons represent their respective systems of values, which play a key role both in the choice of *perspectives* and in the emission of *viewpoints*. $D_1$, $D_2$ and $D_3$ are documents that a mining algorithm has indexed by the topic/tag 'apple'.

Step2: A is a calm person who has time; she browses through $D_1$, $D_2$ and $D_3$ and has a positive feeling about $D_1$ and $D_2$ (she likes both and finds them relevant with respect to 'apple'); the capture of her feedbacks results in linking $D_1$ and $D_2$ to her and reinforcing the links between the documents and the topic 'apple'. B is always in a hurry; he asks the Knowledge Graph the question "which is the shortest path between me and the topic 'apple'?" According to the paths in the diagram, he gets a double answer: B-A-$D_1$-'apple' and B-A-$D_2$-'apple'.

Step3: B's feedback to $D_1$ is positive; this results in reinforcing the path B-A-$D_1$-'apple'. If he would re-ask his question, he would now get only $D_1$.

Step4: C likes to explore; rather than taking a short path she browses through $D_1$, $D_2$ and $D_3$ and has a positive emotion about $D_3$ (she likes it and finds it relevant with respect to 'apple'); this results in linking $D_3$ to her and reinforcing the linking between $D_3$ and the topic 'apple'. At this stage, if A, B and C would ask for the shortest path to 'apple', they would respectively get $D_1$, $D_1$ and $D_3$.

Step5: B is not fully satisfied by $D_1$. He asks for a novel short path between him and the topic 'apple' using a new perspective: he discards the viewpoints expressing his own feelings in order to discover new sources of knowledge. According to this new perspective, B-A-$D_1$-'apple', B-A-$D_2$-'apple'and B-C-$D_3$-'apple' have the same length i.e., $D_1$, $D_2$ and $D_3$ are equidistant from him. B may now discard $D_1$ (already visited) and $D_2$ (already rejected) and read $D_3$.

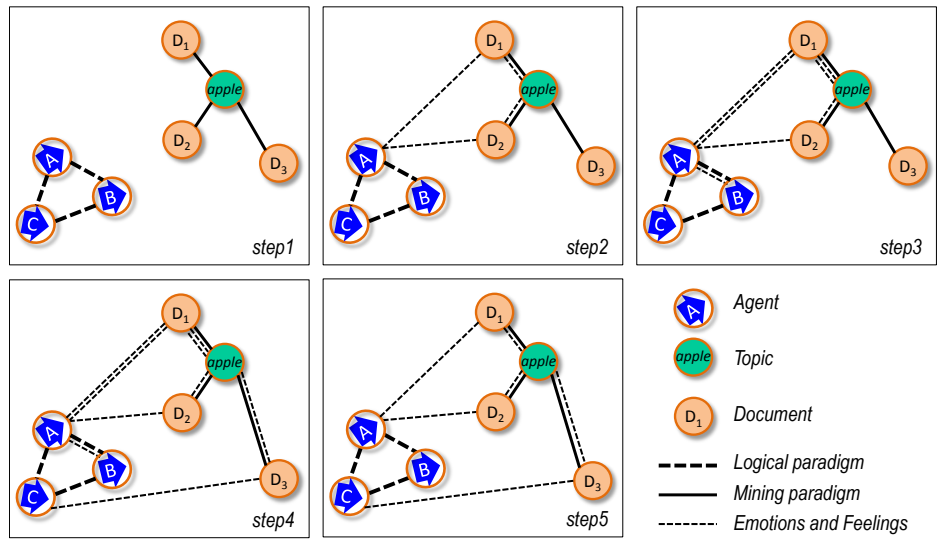

**Fig. 2.** The network of interlinked resources evolves along the attempts of the learners A, B and C to "catch" the topic 'apple' through performing the modules D1, D2 or D. What is figured in the schemas is not the Knowledge Graph itself, but the views (also called Knowledge Maps) resulting from the perspectives; in these maps, the more links between two resources, the closer they are.

Along the five steps of this imaginary case, the evolution of "knowledge paths" follows the metaphor of the selective reinforcement of neural beams, except that this reinforcement is not regulated by a single system of values, rather by a collaboration/competition between the three systems of values of A, B and C. The three co-learners learn as a whole, in a trans-disciplinary way: the dynamics are governed by a topology mixing information and emotions, not by pure logics.

## 5   Conclusion

Starting from the three worlds proposed by K. Popper (the external world of objects and events, the internal world of mind and the world of language), we have browsed through the TNGS of G. M. Edelman and learnt how the learning events are supported by an adaptive neural network, are regulated by our systems of values and occur mainly through social interactions. We have then focused on Web interactions and reformulated the question of the emergence of collective knowledge partially supported by algorithms.

The ViewpointS approach and formalism offer to integrate most if not all these elements in the metaphor of a collective brain; we illustrate through an imaginary case how to trace and enhance collective knowledge acquisition. Within ViewpointS, the three paradigms for knowledge acquisition (logical inferences of the semantic Web, statistical recommendations of the mining community, authority and trust of the social Web) are merged into a knowledge graph of digital traces interpretable by human and artificial agents. Within this graph, the beams of connections are regulated by the individual systems of values that support affect i.e., culture, personality traits, as defined in [22].

What we gain in the integration may be lost with respect to the advantages of each of the three knowledge acquisition paradigms taken individually. For this reason, we are not yet sure if the collective knowledge emerging from ViewpointS graphs and maps (our proposed Collective Brain) will perform competitively with a similar wisdom emerging from each of the three crowds.

Nevertheless: as it has been always the case in the synergies between technological developments and scientific progress, the developments do not ensure scientific discovery, rather may facilitate the process. For instance: Galileo's telescopes did not directly produce the results of modern astronomy, but enabled a significant progress. We hope and believe that our proposed Collective Brain will have a positive impact in understanding and enhancing some aspects of human cognition.